\documentstyle[prl,aps,twocolumn,epsfig]{revtex}

\begin{document}
\twocolumn[\hsize\textwidth\columnwidth\hsize\csname
@twocolumnfalse\endcsname

\title{Heavy Flavor Enhancement as a Signal of Color Deconfinement}

\author{
A.P. Kostyuk$^{a,b}$,
M.I. Gorenstein$^{a,b}$
and 
W. Greiner$^{a}$
}

\address{$^a$ Institut f\"ur Theoretische Physik, Universit\"at  Frankfurt,
Germany}

\address{$^b$ Bogolyubov Institute for Theoretical Physics,
Kyiv, Ukraine}

\date{\today}
\maketitle

\begin{abstract}
We argue that the color deconfinement in heavy ion collisions may 
lead to enhanced production of hadrons with open heavy flavor
(charm or bottom). We estimate the upper bound of this enhancement.
\end{abstract}

\pacs{12.40.Ee, 25.75.-q, 25.75.Dw, 24.85.+p}

]

The production of open heavy flavor (HF) hadrons
(charm and bottom) still remains
a 'terra incognita' of   
heavy ion  physics: neither open charm nor open bottom yields have 
been measured so far. Open charm measurements are only planned
in Pb+Pb collisions at the CERN SPS. 
The standard theoretical picture assumes that the average number 
of hadron pairs with open HF $\langle HF \rangle_{AB(b)}$ created in a 
nucleus-nucleus ($A+B$) collision at given impact parameter $b$
is simply connected\footnote{Here 
we neglect  (anti-)shadowing effects which are expected to
be not very large at SPS and RHIC energies.} 
with the probability to create a HF hadron pair
in a nucleon-nucleon (N+N) collision $\langle HF \rangle_{NN}$:     
\begin{equation}\label{standard}
\langle HF \rangle_{AB(b)}
= N_{coll}^{AB}(b) \langle HF \rangle_{NN}
= N_{coll}^{AB}(b) 
\frac{\sigma_{N N \rightarrow  HF + X}}
{\sigma_{N N}^{inel}}, 
\end{equation}
where $N_{coll}^{AB}(b)$ is the average number of primary 
nucleon collisions, which is determined by the geometry of 
the colliding nuclei, 
$\sigma_{N N \rightarrow  HF + X}$ is the 
total cross section  of the HF hadron pair production in 
$N+N$ collisions and $\sigma_{N N}^{inel}$ is the total inelastic
cross section of $N+N$ interaction.
(Note that in high energy collisions the HF production   
is dominated by the creation of hadrons with open HF.
The HF quarkonia correspond to a tiny fraction 
of the total HF yield and can be safely neglected in our consideration.)

There are however some indirect indications that
an essential deviation from the standard formula 
(\ref{standard}) may exist. 
Recent analysis of the dimuon spectrum 
measured in central Pb+Pb collisions at 158~A~GeV by NA50 
Collaboration \cite{NA50}
reveals a significant enhancement of the dilepton
production in the intermediate mass region (1.5$\div$2.5~GeV) over
the standard sources.
The primary\footnote{Alternative explanations are also suggested 
\cite{alternative}. } 
interpretation attributes this observation to the
enhanced production of open charm \cite{NA50}:
about $3$ times above the direct extrapolation (\ref{standard}) 
from N+N data.
Similar result has been recently obtained in the framework of 
the statistical coalescence model \cite{BMS,coal1,coal2}.  
This model connects the multiplicities of hadrons with open 
and hidden charm. 
It was found \cite{coal1,coal2} that an enhancement of the open charm 
by the factor of about $2 \div 4$ over the direct extrapolation is 
needed to explain the data on the $J/\psi$ multiplicity.
It was suggested in Ref.\cite{coal2} that this enhancement 
may appear due to the broadening of the phase space available for the 
open charm because of the presence of strongly interacting medium.

In the present letter we demonstrate that a deconfined medium
(quark-gluon plasma (QGP) or its precursor) can
make an essential influence on the hadronization of HF (anti-)quarks.
This leads to an enhancement of the HF hadron production in
A+B collisions in
comparison to the direct extrapolation  (\ref{standard}) from 
the N+N data. We restrict ourselves to a rough
estimation of the {\it upper bound}  of possible HF enhancement
due to the color deconfined medium.

The process of production of a HF hadron pair can be 
subdivided into two stages: the hard production
of a HF quark-antiquark pair ($Q\overline{Q}$)
and its subsequent hadronization into observed
particles. Therefore, there is an essential
difference between HF hadron production and, e.g.,
hard dilepton production (the Drell-Yan process):
created $Q$ and $\overline{Q}$ can and even have to interact
with the surrounding quarks and gluons to be transformed into
observed HF hadrons.

To get an intuitive picture of possible medium effects let us 
start from the open HF production in $e^+e^-$ annihilation.
The HF $Q\overline{Q}$ pair created at the first stage, 
hadronizes into observed particles.
The hadronization has a nonperturbative nature.  
Its dynamics  can be qualitatively understood in the framework 
of the string picture. When the distance between $Q$ and 
$\overline{Q}$ reaches the range of the confinement forces, 
a string connected these colored objects is formed.
If the $e^+e^-$  center-of-mass (c.m.) energy $\sqrt{s}$ 
(equal to the invariant
mass of $Q\overline{Q}$ pair $M_{Q\overline{Q}}$)
lies well above the corresponding HF meson threshold $2m_M$  
(equal to $2m_D$ or $2m_B$ for $c\overline{c}$ and
$b\overline{b}$ quarks, respectively), $Q$ and $\overline{Q}$ break 
the string into two (or more) peaces, so that the final state 
contains a HF hadron pair (and possibly a number of light hadrons).
However, when the $e^+e^-$ c.m. energy 
exceeds the heavy quark threshold 
($\sqrt{s} > 2 m_Q$) but 
lies below the corresponding HF meson threshold $2m_M$
($\sqrt{s} < 2 m_M$), the string cannot be broken 
and the open HF hadron pair  {\it can not} be formed. 

Let us imagine now the  $e^+e^-$ annihilation inside
a deconfined medium.
Due to the Debye screening, no string is formed between colored 
objects in this case. If the heavy $Q$ and $\overline{Q}$ are created,
they {\it can} fly apart within the medium as 
if they were free particles. It does not matter whether their
initial invariant mass $M_{Q\overline{Q}}$ exceeds the 
corresponding hadron threshold or not. 
The created $Q\overline{Q}$ pair {\it will be able} to form  
a HF hadron pair at the stage of QGP hadronization.
This means that {\it 
the $e^+e^-$ annihilation inside the QGP would
produce HF hadrons even if the collision energy is not sufficient
for producing these hadrons in the vacuum.  
}

In N+N or A+B collisions the HF $Q\overline{Q}$ pairs 
are produced due to hard {\it parton} interactions.
The calculations in the leading order of the perturbative quantum
chromodynamics (pQCD) show  that 
a great fraction of $Q\overline{Q}$ pairs are created 
with invariant masses $M_{Q\overline{Q}}$ below the 
corresponding meson threshold $2m_M$ 
even at the largest RHIC energy.
If this $Q\overline{Q}$ pair creation takes place in the deconfined
medium, which is expected to be formed in high energy A+B collisions, 
the presence of such a medium makes possible a hadronization of  
these pairs. {\it  This should lead to an enhancement of the 
HF hadron production in A+B collisions in comparison to the
standard result (\ref{standard}) obtained within the 
direct extrapolation of the N+N data.}

There are of course essential differences between the open HF
hadron production
in the $e^+e^-$ annihilation and in N+N or A+B collisions.
Even in N+N collisions, when 
no deconfined medium is expected, the created $Q\overline{Q}$ pair 
{\it can interact} 
with the spectator partons and has 
therefore a chance to form HF hadron pair even 
if its primary invariant mass  was insufficient 
for this process. Moreover, in contrast to the $e^+e^-$ annihilation, 
the most of $Q\overline{Q}$ pairs are created  
in the color octet state and therefore they {\it have to interact}
with the spectators to form a color neutral final state. 
Instead of  
breaking the string, the $Q$ and $\overline{Q}$ can form hadron states
by means of coalescence with light spectator  (anti-)quarks\footnote{
This mechanism is responsible, e.g., for the reaction 
$pp \rightarrow p \Lambda_c \overline{D}$.}. 

As no theoretical descriptions of this complicated process
exist, we restrict ourselves to a rough estimation of 
the {\it upper bound}  of possible HF hadron enhancement
due to the color deconfined medium.  We {\it assume} that 
\begin{itemize}
\item 
In the case of N+N collisions,  {\it no}  subthreshold  
$Q\overline{Q}$ pairs contribute to the HF hadron production\footnote{
In other words, we assume that the  
interaction with the spectators does not change 
the energy of 
the  $Q\overline{Q}$ pair and no coalescence with the spectator
(anti-)quarks takes place.}. 
%
\item 
In the case of A+B collisions, provided that the 
deconfined medium is formed, {\it all} $Q\overline{Q}$ pairs
hadronize into particles with the open HF\footnote{A small   
fraction of them  form also the hidden heavy-flavor mesons, 
but it can be safely neglected.}.  
\end{itemize}
 
\noindent
The first assumption looks
reasonable at low collision energies, whereas
to justify the second one high energies are evidently preferable.
This means that assuming validity of the both 
statements we overestimate the expected HF enhancement effect and,
therefore, the above assumptions give its {\it upper bound}.

\vspace{0.3cm}       
We make now the numerical estimates which follow from the above
assumptions.
The total cross section of  heavy $Q \overline{Q}$ pair production by 
colliding nucleons  is given by the formula 
(see e.g. Ref.\cite{Combridge}) 
\begin{eqnarray}\label{sigma_tot}
&& \sigma_{N N \rightarrow Q  \overline{Q} + X}(s) \  = \  \\ 
&& \sum_{(1,2)}
 \int_0^1 dx_1 \int_0^1 dx_2  f_1(x_1,\mu_F)   f_2(x_2,\mu_F) 
\hat{\sigma}_{1   2 \rightarrow Q   \overline{Q} +\hat{X} }(\hat{s})  \  ,
\nonumber 
\end{eqnarray}
where $s$ is the squared c.m. energy of the colliding 
nucleons, $x_1$ ($x_2$) is the fraction of the momentum of the first
(second) nucleon carried by the parton $1$ ($2$),  $f_1$ and $f_2$ 
are the fractional-momentum distribution functions or structure 
functions, $\mu_F$ is the factorization scale,
$\hat{\sigma}_{1 2 \rightarrow Q \overline{Q}}(\hat{s})$ is the cross 
section of heavy quark-antiquark pair production by interacting
partons at squared center-of-mass energy  $\hat{s}$. For 
ultrarelativistic nucleons, $\hat{s}$ is given by the formula 
$\hat{s}=x_1 x_2 s$. The sum in the right hand side of 
Eq.(\ref{sigma_tot}) runs over all the pairs of parton types, that 
give nonzero contribution to the production cross section.

We restrict ourselves to the leading order of pQCD. 
In this case, two basic processes of  heavy flavor creation 
have to be taken into account: the
gluon fusion $gg \rightarrow Q\overline{Q}$  and the light quark-antiquark 
annihilation $ q \overline{q} \rightarrow Q\overline{Q} $. So the sum in  
Eq.(\ref{sigma_tot}) includes 
$(1,2)\ =\ (g,g),\  (\overline{q},q),\  (q,\overline{q}) $,
where q in its turn runs over the light flavors   $q=u,\  d,\  s$.
The corresponding parton cross sections are given by the 
formulas \cite{Combridge}:
\begin{eqnarray}
&& \hat{\sigma}_{g g \rightarrow Q  \overline{Q}}(\hat{s})  =  
\frac{\pi \alpha^2(\mu_R)}{3 \hat{s}} \\
&& \times \left [
- \left(
7 + \frac{31 m_Q^2}{\hat{s}}
\right) \frac{1}{4} \chi
+  \left(
1 +  \frac{4  m_Q^2}{\hat{s}} +  \frac{ m_Q^4}{\hat{s}^2}
\right) \log \frac{1+\chi}{1-\chi}
\right] \nonumber
\end{eqnarray}       
and 
\begin{equation}
\hat{\sigma}_{q \overline{q} \rightarrow Q  \overline{Q}}(\hat{s})  = 
\frac{8 \pi \alpha^2(\mu_R)}{27 \hat{s}}
\left(
1 +  \frac{2  m_Q^2}{\hat{s}} 
\right)  \chi  \  ,
\end{equation}
where  $\chi = \sqrt{1 -  4  m_Q^2/\hat{s} }\ $,
$\mu_R$ is the renormalization scale and
$m_Q$ is the mass of the heavy quark. The masses of light quarks 
are neglected.

Eq.(\ref{sigma_tot}) can be rewritten in the form 
\begin{equation}\label{sigma_tota}
\sigma_{N N \rightarrow Q  \overline{Q} + X}(s) \  = \ 
 \int_{(2m_Q)^2}^s d\hat{s}~ 
\frac{d  \sigma_{N N \rightarrow Q  \overline{Q} + X}}{d \hat{s}}   \   , 
\end{equation}
where the differential cross section with respect to the 
squared invariant mass $\hat{s}=M_{Q\overline{Q}}^2$ of the 
$Q\overline{Q}$ pair is given by the formula
\begin{eqnarray}\label{dsigma}
&&\frac{d  \sigma_{N N \rightarrow Q  \overline{Q} + X}}{d \hat{s}}   = \\ 
&& \frac{1}{s} \sum_{(1,2)}
\hat{\sigma}_{1   2 \rightarrow Q   \overline{Q} +\hat{X} }(\hat{s}) 
 \int_{ \hat{s}/s - 1}^{ 1-\hat{s}/s } 
dx_L  \frac{ f_1(x_1,\mu_F)   f_2(x_2,\mu_F)}{x_1 + x_2} 
  \  , \nonumber
\end{eqnarray}
where
\begin{eqnarray}
x_1 &=& \sqrt{\left(\frac{x_L}{2}\right)^2  + \frac{\hat{s}}{s} } + 
\frac{x_L}{2} \label{x1}\\ 
x_2 &=& \sqrt{\left(\frac{x_L}{2}\right)^2  + \frac{\hat{s}}{s} } - 
\frac{x_L}{2} \label{x2}.
\end{eqnarray}

\begin{figure}[p]
\begin{center}
\vfill
\epsfig{file=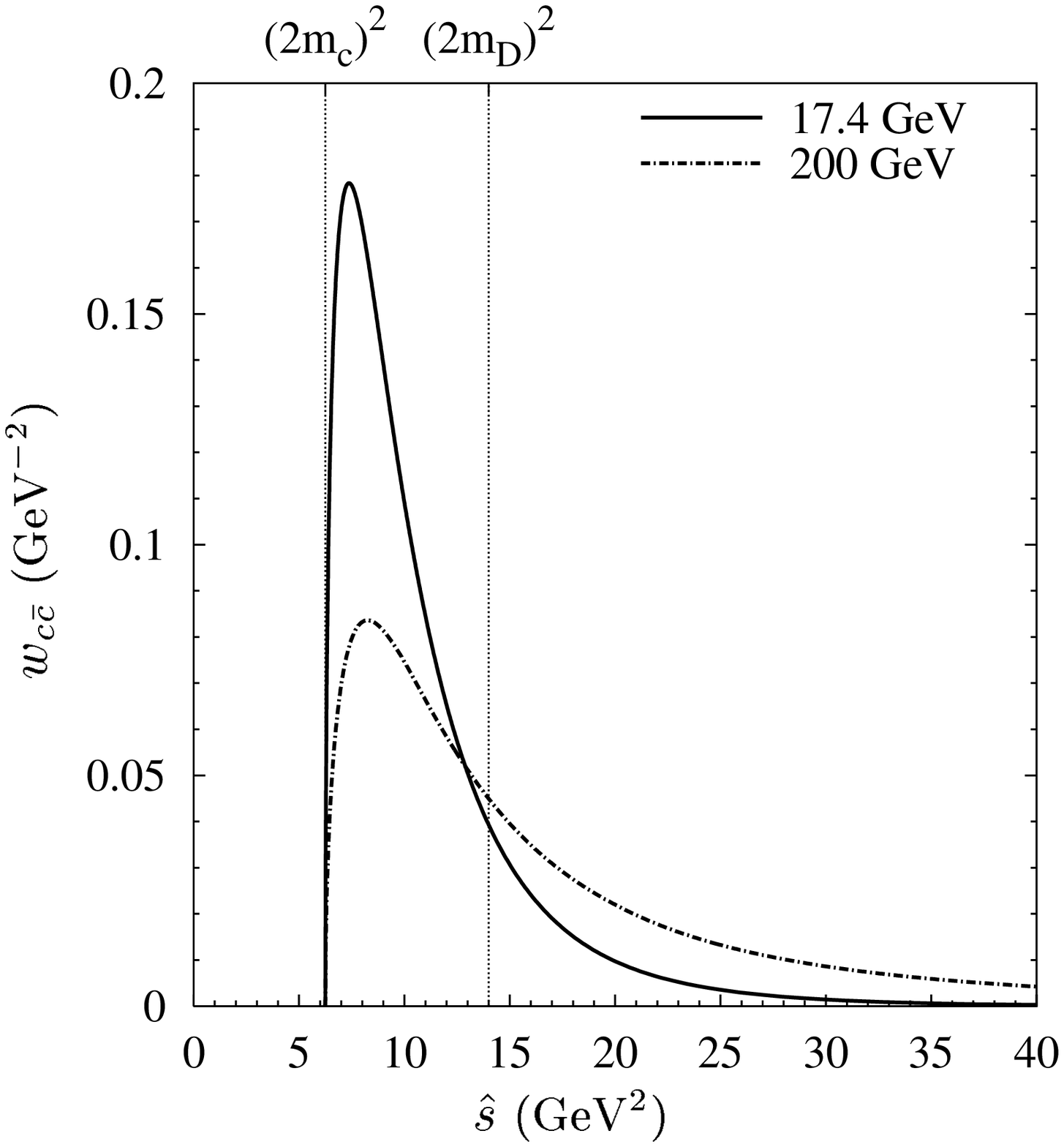,height=18cm}
\mbox{}\\
\vfill
\caption{The distribution of $c\overline{c}$ pairs created in nucleon-nucleon
collisions versus their squared invariant mass $\hat{s}$. 
The two curves correspond to different c.m. energies of the
colliding nucleons: SPS energy $\sqrt{s}=17.4$~GeV and maximum RHIC 
energy $\sqrt{s}=200$~GeV. Great part of the $c\overline{c}$ pairs have the
invariant mass below the $D$-meson threshold $2 m_D$.   
\label{w_c}
}
\end{center}
\end{figure}

\begin{figure}[p]
\begin{center}
\vfill
\epsfig{file=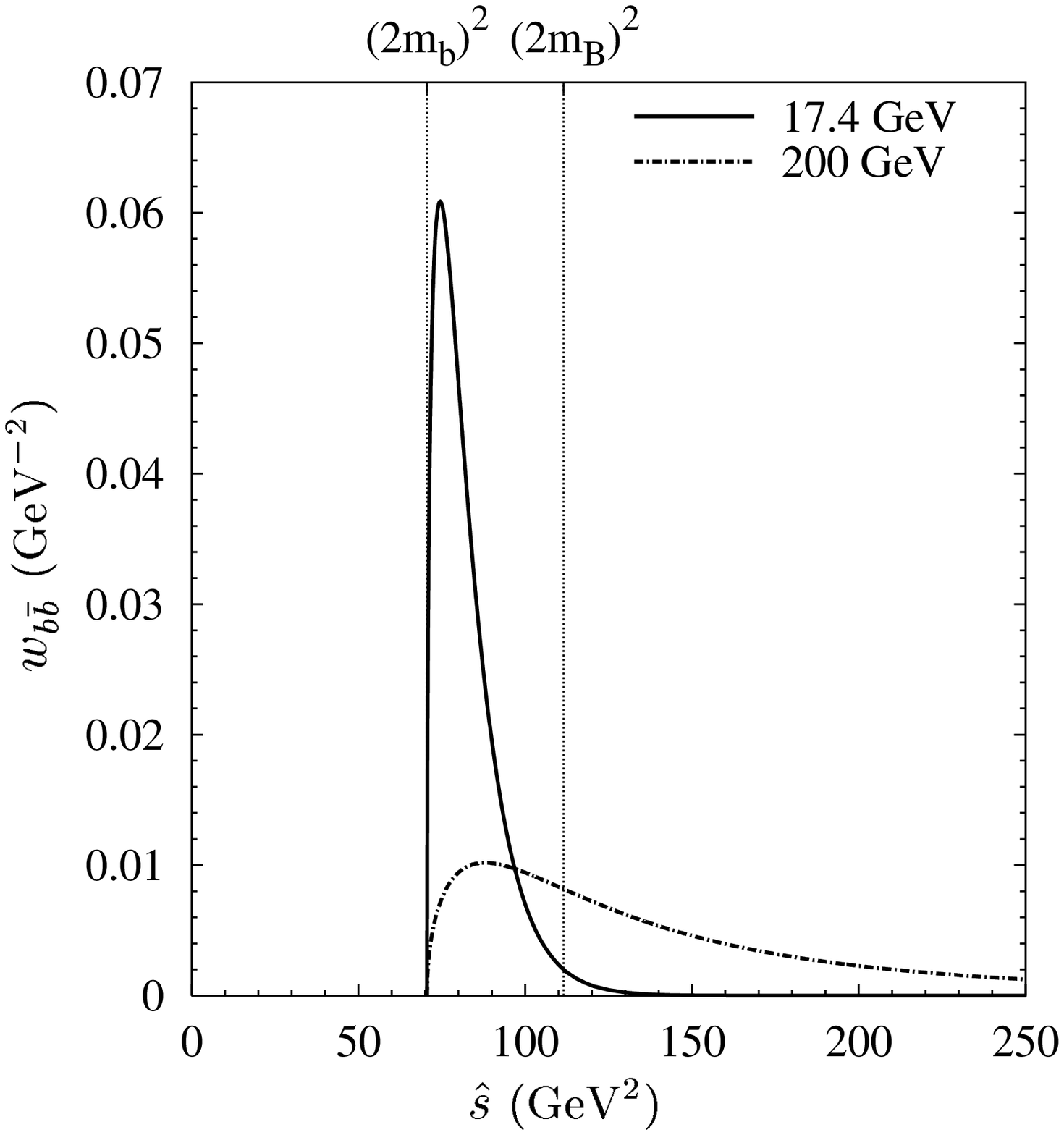,height=18cm}
\mbox{}\\
\vfill
\caption{The same as in Fig.\ref{w_c} but for $b\overline{b}$ pairs.
\label{w_b}
}
\end{center}
\end{figure}

The probability distributions of $Q\overline{Q}$ pairs with respect to
$\hat{s}$ 
\begin{equation}
w_{Q \overline{Q}}(\hat{s}; s) = 
\frac{d  \sigma_{NN \rightarrow Q  \overline{Q} + X}/d \hat{s}}
{\sigma_{NN \rightarrow Q  \overline{Q} + X}(s)} 
\end{equation}
are shown in Fig. \ref{w_c}  and Fig. \ref{w_b} for charm and bottom, 
respectively. The computation were done using the CERN library
of parton distribution functions PDFLIB \cite{PDFLIB}. The default 
set of structure functions MRS (G) \cite{MRS-G} was chosen.
The HF quark masses are fixed as $m_c=1.25$~GeV for charm
and $m_b=4.2$~GeV for bottom, the c.m. energy of the
colliding 
parton pair was used as the renormalization and factorization scales: 
$\mu_F=\mu_R=\sqrt{\hat{s}}$. 

We estimate now the upper bound of the HF enhancement in 
A+B collisions. We assume that 
in N+N collisions the HF $Q\overline{Q}$ pairs cannot hadronize, unless
its c.m.
energy exceeds the corresponding HF hadron threshold.
Therefore, to calculate the total HF hadron production 
cross section we cut the integral in Eq.(\ref{sigma_tota}) at its
lower bound by the corresponding meson threshold\footnote{  
One has to cut the
integral also at its upper bound, 
to respect the baryonic number conservation.  Our 
calculations, however, show that this cut is important 
only at very low collision energies: $\sqrt{s} \cong 2 m_N + 2 m_M$.}:
\begin{equation}\label{sigma_vac}
\sigma_{N N \rightarrow  HF + X} 
= \ \int_{(2m_M)^2}^{(\sqrt{s} - 2 m_N)^2} d\hat{s} 
\frac{d  \sigma_{N  N \rightarrow Q  \overline{Q} + X}}{d \hat{s}}   \   , 
\nonumber
\end{equation}
where $m_M$ is the mass of the lightest meson containing corresponding 
HF quark ($D$-meson for the charm and 
$B$-meson for the bottom), $m_N$ is the nucleon mass. 

In contrast, when two nucleons interact in the deconfined medium
(as in high energy A+B collision),
our assumption states that all $Q\overline{Q}$ pairs survive
and form HF hadrons at the stage of  the QGP hadronization.
Therefore, the cross section $\sigma_{N N \rightarrow  HF + X}$ in the
formula (\ref{standard}) should be replaced by the 
cross section $\sigma_{N N \rightarrow Q  \overline{Q} + X}$. Hence for the 
upper bound of the enhancement factor we use the formula
\begin{equation}\label{enhancement}
E^{\mbox{\footnotesize max}}(s) = 
\frac{\sigma_{N N \rightarrow Q  \overline{Q} + X}(s)}
{\sigma_{N N \rightarrow HF + X}(s)}.
\end{equation}

\begin{figure}[p]
\begin{center}
\vfill
\epsfig{file=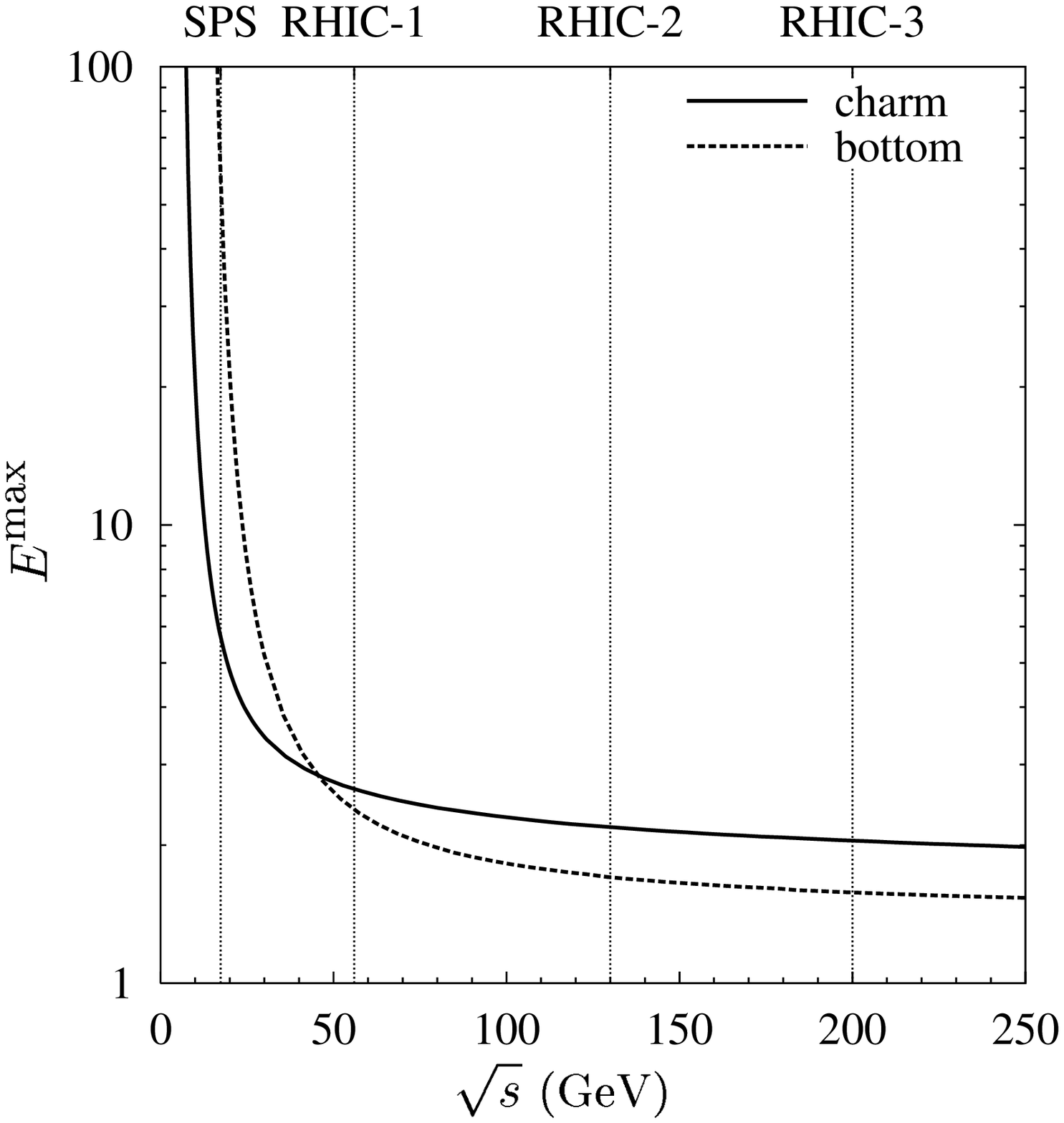,height=18cm}
\mbox{}\\
\vfill
\caption{The upper bound of heavy flavor enhancement versus the
c.m. energy of the colliding nucleons. The vertical lines
show SPS and RHIC energies. 
\label{Emax}
}
\end{center}
\end{figure}

The behavior of $E^{\mbox{\footnotesize max}}(s)$ 
for charm and bottom is shown 
in Fig.\ref{Emax}. It is seen that the largest effect is expected 
at low energies. Therefore an experimental study of the effect 
should be done at the minimum energy, where the deconfinement medium 
is expected to be formed and the inclusive cross-section of HF
production is large enough to make its measurement feasible.

The upper bound of open charm enhancement at SPS energy is by the 
factor of about $5 \div 6$. This means that the enhanced 
production of open charm hadrons by the factor $2 \div 4$ found in
Ref.\cite{NA50} and Refs.\cite{coal1,coal2} can be explained 
by the influence of the deconfined medium.

We conclude that the deconfined medium, which is expected to be 
formed in nucleus-nucleus collisions can influence the process of 
hadronization of heavy quarks, this leads to the enhanced production 
of hadrons with open heavy flavors (charm and bottom). The rough 
estimation of the upper bound of the 
effect at SPS energies is found to be large enough to explain the 
indirect experimental data \cite{NA50} and the phenomenological  
evaluations \cite{coal1,coal2}. We consider the enhancement of the
heavy flavor yield as a possible signal of the color deconfinement.
The direct measurement of open charm and open bottom in nucleus-nucleus 
collisions could be important for the confirmation of the 
quark-gluon plasma formation.

\acknowledgments

The authors
are thankful to K.A.~Bugaev, M.~Ga\'z\-dzicki, L.~McLerran,
I.N.~Mishus\-tin, G.C.~Nayak,  K.~Red\-lich, E.V.~Shu\-ryak and 
H.~St\"ocker for
fruitful discussions and their interest to the work.
We appreciate criticism and useful comments of L.~Frankfurt and 
L.~Gerland.
The financial support of DAAD, Germany is acknowledged.
The research described in this publication was made possible in part by
Award \# UP1-2119 of the U.S. Civilian Research and Development
Foundation for the Independent States of the Former Soviet Union
(CRDF).

\pagebreak
\mbox{}
\pagebreak
\widetext

\pagebreak
\mbox{}
\pagebreak

\widetext

\pagebreak
\mbox{}
\pagebreak

\begin{thebibliography}{99}

\bibitem{NA50}
M.~C.~Abreu {\it et al.}  [NA38 Collaboration],
Eur.\ Phys.\ J.\ {\bf C14}, 443 (2000).

\bibitem{alternative}
C.~Spieles {\it et al.},
Nucl.\ Phys.\  {\bf A638}, 507 (1998);
R.~Rapp and E.~Shuryak,
Phys.\ Lett.\  {\bf B473}, 13 (2000);
K.~Gallmeister, B.~K\"ampfer and O.~P.~Pavlenko,
Phys.\ Lett.\  {\bf B473}, 20 (2000).

\bibitem{BMS}
P.~Braun-Munzinger and J.~Stachel,
Phys.\ Lett.\ {\bf B490}, 196 (2000)
[nucl-th/0007059].

\bibitem{coal1}
M.~I.~Gorenstein, A.~P.~Kostyuk, H.~Stocker and W.~Greiner,
hep-ph/0010148.

\bibitem{coal2}
M.~I.~Gorenstein, A.~P.~Kostyuk, H.~Stocker and W.~Greiner,
hep-ph/0012015.


\bibitem{Combridge}
B.~L.~Combridge,
Nucl.\ Phys.\ {\bf B151}, 429 (1979).

\bibitem{PDFLIB}
H.~Plothow-Besch,
calculations,''
Comput.\ Phys.\ Commun.\ {\bf 75}, 396 (1993);
H.~Plothow-Besch,
Int.\ J.\ Mod.\ Phys.\ {\bf A10}, 2901 (1995).

\bibitem{MRS-G}
A.~D.~Martin, W.~J.~Stirling and R.~G.~Roberts,
Phys.\ Lett.\ {\bf B354}, 155 (1995)
[hep-ph/9502336].

\end{thebibliography}
\end{document}